\documentclass[journal, 10pt]{IEEEtran}

\usepackage{amsmath}
\usepackage{amsfonts}
\usepackage{amssymb}
\usepackage{amsthm}
\usepackage{color}
\usepackage{array}
\usepackage{cite}
\usepackage{enumerate}
\usepackage{graphicx}
\usepackage[hang]{subfigure}
\usepackage{epstopdf}
\usepackage{epsfig}
\usepackage{footmisc}
\usepackage{amsbsy}
\usepackage{bm}
\usepackage{cases}
\usepackage{multirow}
\usepackage{algorithmic}
\usepackage{comment}
\usepackage{times}
\usepackage{paralist}
\usepackage[ruled]{algorithm2e}
\usepackage{setspace}
\usepackage[breakable]{tcolorbox}
\IEEEoverridecommandlockouts

\begin{document}
\title{Semantic-Effectiveness Filtering and Control for \\ Post-5G Wireless Connectivity}

\author{Petar Popovski~\IEEEmembership{Fellow,~IEEE,}
        Osvaldo Simeone~\IEEEmembership{Fellow,~IEEE,}
        Federico Boccardi~\IEEEmembership{Senior Member,~IEEE,} \\
        Deniz G\"und\"uz~\IEEEmembership{Senior Member,~IEEE,}
        and Onur Sahin~\IEEEmembership{Member,~IEEE}
\thanks{The first two authors are in alphabetical order, as well as the last three authors. Petar Popovski (petarp@es.aau.dk) is with the Department of Electronic Systems, Aalborg University, Aalborg, Denmark. Osvaldo Simeone (osvaldo.simeone@kcl.ac.uk) is with the Centre for Telecommunications Research, King’s College London, London, United Kingdom. Federico Boccardi (federico.boccardi@ieee.org) is a Visiting Professor at University at Bristol, United Kingdom. Deniz G\"und\"uz (d.gunduz@imperial.ac.uk) is with the Imperial College London, United Kingdom. Onur Sahin (Onur.Sahin@InterDigital.com) is with InterDigital, London, United Kingdom. This work has received funding from the European Research Council (ERC) under the European Union Horizon 2020 research and innovation program (grant agreements 725731, 677854 and 648382).}}
\maketitle

\begin{abstract}
The traditional role of a communication engineer is to address the technical problem of transporting bits reliably over a noisy channel. With the emergence of 5G, and the availability of a variety of competing and coexisting wireless systems, wireless connectivity is becoming a commodity. This article argues that communication engineers in the post-5G era should extend the scope of their activity in terms of design objectives and constraints beyond connectivity to encompass the semantics of the transferred bits within the given applications and use cases. To provide a platform for semantic-aware connectivity solutions, this paper introduces the concept of a \textit{semantic-effectiveness (SE) plane} as a core part of future communication architectures. The SE plane augments the protocol stack by providing standardized interfaces that enable information filtering and direct control of functionalities at all layers of the protocol stack. The advantages of the SE plane are described in the perspective of recent developments in 5G, and illustrated through a number of example applications. The introduction of a SE plane may help replacing the current ``next-G paradigm'' in wireless evolution with a framework based on continuous improvements and extensions of the systems and standards.  
\end{abstract}


% \fi
\IEEEpeerreviewmaketitle

%\section{Introduction}

%\begin{quote} \emph{You're talking a lot, but you're not saying anything}\\
%\emph{When I have nothing to say, my lips are sealed} \\
%\emph{Say something once, why say it again?} \\
%Psycho Killer, Talking Heads \end{quote}

\section{Introduction}

\subsection{Motivation}
Since the early days of communication systems engineering, the complexity of the communication process has motivated a compartmentalization of the subject into separate disciplines. Shannon and Weaver famously identified three levels of problems within the broad subject of communication~\cite{shannon1949mathematical}: 
\begin{description}
\item[A.] \emph{Technical problem:} The symbols conveying information should be reliably transmitted to the recipient;
\item[B.] \emph{Semantic problem:} The meaning conveyed by the transmitted symbols should accurately reflect the intentions of the sender;
\item[C.] \emph{Effectiveness problem:} The conduct or action of the system in response to communications should be effective in accomplishing a desired task.  
\end{description}
We illustrate these three levels through two  contemporary examples. In online shopping, the technical problem amounts to the establishment of a reliable end-to-end connection; the semantic problem to the definition of the most informative and appealing display of products on the webpage; and the effectiveness problem relates to the rate of success in terms of sales figures or targeted advertising. In a remote monitoring application via an Internet-of-things (IoT) system, the technical problem is the transmission of collected data from the IoT sensors to a cloud processor; the semantic problem is the extraction of knowledge from data through data analytics; and the effectiveness problem concerns the impact of the action taken by a human or by relevant actuators in response to the data collected from the sensors.

Traditionally, communication engineers have been solely concerned with the technical problem. This focus has been arguably one of the key reasons for the unprecedented success of the communication technology in the last decades. However, as communication technologies have matured into the status of a commodity, the barycenter of technology and business has migrated in the direction of the semantic and the effectiveness problems. The best indicator for this trend is the economic, as well as political clout, of the most valuable digital tech companies -- Facebook, Apple, Amazon, Netflix, Google (FAANG) -- that are built to address these problems rather than the technical problem. Confining their activities to the technical problem may cause today's communication engineers to be stuck in what we may refer to as ``Shannon's trap''. The ongoing migration of communication engineers towards methods, or even job positions, related to machine learning (ML) and artificial intelligence (AI) provides a tangible evidence of this trend. 

% Shannon himself appears to be well-aware of this danger, as he diversified his later research to encompass a variety of engineering activities beyond the technical problem, most notably by means of work around artificial intelligence (AI). 
% \color{red} [D: this last sentence seems a bit too speculative to me. Can we give evidence that Shannon actually looked into the semantic or effectiveness problems? his AI work --as far as I know-- does not deal with communication per se.]\color{black} 

\begin{figure}[!t]
\centering
\includegraphics{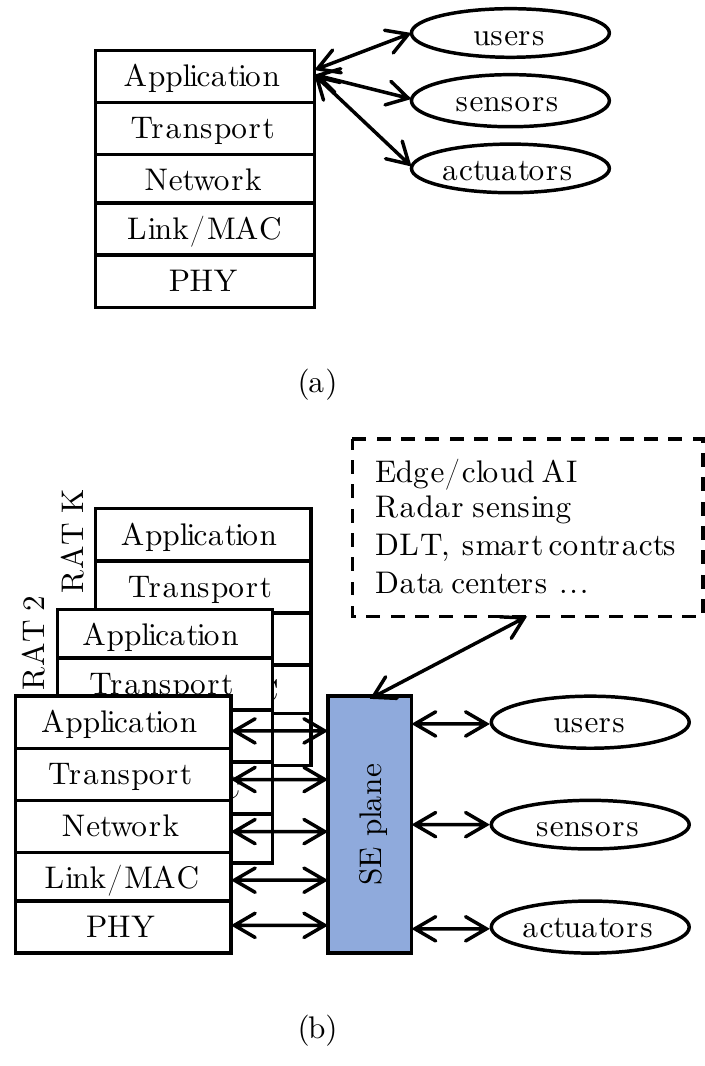}%
\caption{Interaction of the users, sensors, and actuators with the protocol stack. (a) 2G-5G cellular networks; (b) proposed architectural evolution in the post-5G wireless systems that includes the SE plane. The figure emphasizes the fact that the SE plane can generally interact and coordinate multiple Radio Access Technologies (RATs), rather than being dedicated to a single cellular interface.}
\label{fig1}
\end{figure}

\subsection{An Augmented Protocol Architecture}

The key thesis of this article is that, while designing solutions for post-5G wireless connectivity, wireless communication engineers should expand their efforts beyond the technical problem to address the semantic and effectiveness problems. In light of the above trends, as illustrated in Fig.~\ref{fig1}, we propose to move away from the current protocol architecture whereby the interfaces between users, sensors, and actuators carrying information relevant to semantic and effectiveness problems are limited to the application layer. In contrast, in the proposed architecture, the protocol stack is augmented with a plane that exposes Application Programming Interfaces (APIs) between, on the one end, users, sensors, and actuators, and, on the other end, all layers of the Radio Access Network (RAN) protocol stack. 
%\OS{Are we talking only about the RAN?} 
The APIs enable the extraction of information to be processed via data analytics tools and the direct control of functionalities at any layer. This in turns allows the standardized integration in the communication protocol of data-driven, i.e., ML/AI, methods at a level that is within the purview of the wireless communication engineer.

%\OS{Should we also highlight data and control planes? What is the relationship between data and control planes and semantic plane? I think that all this should be clarified}

As we will discuss, a related effort has been recently made by the 3rd Generation Partnership Project (3GPP) at the level of the 5G core network. With Release 16, 3GPP has in fact introduced standardized interfaces between network functions within the control plane of the core network. These interfaces carry data on protocol operations and performance metrics that can be used for data analytics and resource management. On the one hand, the proposed architecture extends the idea of defining standardized interfaces to the RAN protocol stack, and, on the other, it allows not only for \emph{information extraction} but also for direct \emph{control}. We refer to this new plane -- whose main desired properties will be described here along with some use cases -- as \emph{semantic-effectiveness (SE) plane}. The SE plane covers functionalities concerning both semantic and effectiveness problems, since the two aspects are often intertwined in terms of both data collection and control. 

The rest of the paper is organized as follows. In Sec. II, we provide context and discuss basic functionalities of the SE plane. Sec. III describes a number of exemplifying use cases. In Sec. IV, we place the proposal in the context of current standarization efforts within 3GPP.  Finally, Sec. V concludes the paper by covering key implementation challenges and by offering an outlook.

% \color{red} [D: in this section we emphasize these two problems, semantic and effectiveness, and give explicit examples, but in the rest of the paper we do not distinguish between the two, and indeed never mention effectiveness again. From what I understand the semantic level, even though we call it semantic, also addressed the effectiveness problem -- maybe even more so. I suggest we explicitly state that the semantic plan addresses both problems. ]\color{black} 

\section{SE plane: Context and Overall Functionality}
In this section, we first provide some context to further motivate the proposed augmented protocol stack illustrated in Fig. 1, and then we describe the main properties and functionalities of the proposed SE plane. 
\subsection{Connectivity Trends}

In order to frame the proposal in the proper technological context, we briefly summarize here some relevant current trends in wireless connectivity. We will return to some of the main points in later sections of the paper.

    \noindent $\bullet$ In the post-5G era, systems for wireless connectivity will be abundant and mature, shifting the performance bottleneck -- and, with it, the focus of research -- to layers and services that leverage physical-layer connectivity primitives;  
    %\OS{we wrote core network -- but isn't our proposal for the RAN?}. %While this does not imply that no further advances are expected in physical-layer technology, it is envisaged there will be more substantial and conceptual developments in 
    
   \noindent $\bullet$ Recent advances in ML, also popularly referred to as AI, will enable the extraction of information by means of pattern recognition within complex data streams at all layers of the protocol stack (see, e.g., \cite{simeone2018very});
   
    \noindent $\bullet$ There is an increased demand for wireless services with ultra-low latency and real-time interactive communications, for use cases such as Virtual Reality (VR), Augmented Reality (AR), and remotely controlled robots and drones. The need for ultra-low latency will increase the focus on predictive communications and networking based on ML techniques.
    
    \noindent $\bullet$ As opposed to classical latency, in an increasing number of applications, a more relevant metric is the Age of Information (AoI), which captures the freshness of information by taking into account the dynamics, and hence semantics, of the information source; \cite{ForeverYoung}; %\PP{PP: Deniz, I would not put AoI as a separate bullet, but as part of latency. }
    
    \noindent $\bullet$ The architecture of 5G systems, and very likely of post-5G networks, is evolving away from mere softwarization and virtualization of well-defined Network Functions (NFs)
    towards a Service-Based Architecture (SBA). SBA will enable self-contained and reusable NFs to exchange data via efficient Service-Based Interfaces (SBIs), hence enabling data-driven interactions among NFs \cite{3GPPTS23.288};
    
    %(e.g., $NF_{A}\leftrightarrow NF_{B}$) \PP{Not sure what this notation means, please elaborate, Onur or Fede.} 
    
  \noindent $\bullet$ With the convergence of heterogeneous data- and task-oriented services over the cellular infrastructure, data streams will contain a large fraction of \emph{semantic or effectiveness redundancy}, that is, of data that is delivered to the application layer, but ends up not being relevant or useful. A notable example is the transmission of data for the training of ML models whose source cannot be trusted -- a semantic problem -- or whose inclusion does not lead to significant updates in the current model -- an effectiveness problem;
  
  %This issue is a direct consequence of Shannon's positioning of the communication engineering, according to which the network is largely oblivious to the semantics of the information bits being transported~\cite{shannon1949mathematical};
  
   \noindent $\bullet$ 5G networks will be characterized by an increased \emph{protocol overhead} in order to ensure security, privacy, provenance, as well as trust through a Distributed Ledger Technology (DLT)~\cite{christidis2016blockchains}. This protocol overhead will contain potentially significant semantic and/or effectiveness redundancy. For example, the provenance of data that is no longer relevant for an application of interest need not be validated using, e.g., access to DLT.

\subsection{SE Plane: Basic Functionalities}

\begin{figure}[!t]
\centering
\includegraphics{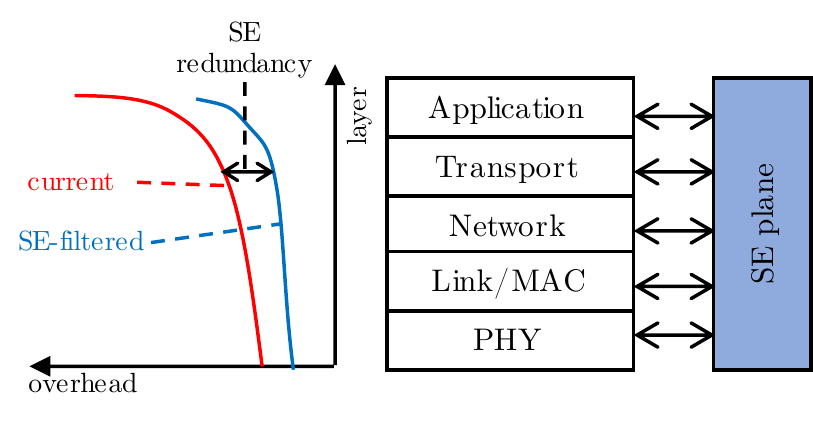}%
\caption{The role of SE filtering in decreasing both application and protocol overheads.}
\label{fig2}
\end{figure}

In order to facilitate integrated solutions to the technical, semantic, and effectiveness problems, we propose the introduction of a \emph{SE plane} in the communication protocol architecture as illustrated in Fig.~\ref{fig1}. As we will detail, the SE plane provides standardized APIs that enable information filtering and direct control at all layers of the protocol stack based on semantic and effectiveness metrics. 

%\OS{I think that this section should be expanded and that we should comment on Fig. 2. I would also modify Fig. 2 to exclude the "layers" and by also illustrating semantic control}

Through its standardized interfaces to all layers of the protocol stack, the SE plane makes it possible to:
\begin{itemize}
\item\emph{Filter} information packets that are \emph{irrelevant}, which is a semantic problem,  or \emph{not useful}, which is an effectiveness problem, by cross-checking against the state of the ML/AI models or trusted DLT tools such as blockchains and smart-contract ledgers. The effect of SE filtering is illustrated on Fig.~\ref{fig2}: by removing SE redundancy, SE filtering aims at reducing overhead at all layers of the protocol stack;
\item \emph{Control} the operation of the communication protocol by providing a direct interface between communication functionalities, at one end, and actuators and sensors, at the other. SE control at all layers of the protocol stack can enable a more efficient use of communication and computing resources, and it opens the way for a principled and controlled introduction of ML/AI tools and techniques into the design of the wireless communication protocol. As discussed in the examples in the next section, SE control can enhance both semantic and effectiveness performance metrics.
\end{itemize}

As we review in Sec. IV, the current standardization efforts have a limited scope in terms of the type of information that can be accessed by ML/AI tools. In particular, the standardized interface only concerns the core network, and it is one-directional, from network functions (NFs) in the core network to cognitive functionalities based on ML/AI. As a result, innovative ``AI'' solutions relying on network analytics are constrained to take place solely within the core networks and within the network elements of a given operator. The latter constraint stems from the absence of standardized control interfaces that are directed from the SE plane to the NFs. The SE plane outlined here addresses both limitations by extending the scope of the standardized APIs to the RAN and by allowing also for direct control of communication functionalities.

At a high level, the functionalities introduced for the SE plane pave the way for the evolution of communication systems according to the following principles and objectives:
\begin{enumerate}
\item to provide new use cases for users and businesses;

\item to embrace ML with the aim of transforming communication networks into adaptive and predictive distributed systems; 

\item to open the value chain in the direction of ``open radio'' with re-defined standardization models and releases. 

\item to ensure security, privacy and trust by allowing a better control of the data and its use.
\end{enumerate}

In the next section, we cover a number of specific examples and use cases in order to illustrate applications of the SE plane.

\section{Implementing the SE plane}

While the specification of the detailed functional blocks and operation of the SE plane is an elaborate task that is outside of the scope for this paper, this section provides several instances that illustrate the functionality and the benefits from the SE plane beyond the current state of the art. 

\subsection{SE Filtering for Immersive and Tactile Applications}

Consider an augmented reality (AR) application, in which collocated users have both physical and digital interactions, e.g., within social networks or video games. Broadband and low-latency traffic packets are required to be carried over wireless connections, since videos and audio signals are complemented by haptic and other sensory signals for an immersive experience. Low-latency transmissions require almost instant access to radio resources. This can be ensured by reserving dedicated resources for low-latency transmissions. This is illustrated through the simple Medium Access Control (MAC) frame in Fig.~\ref{fig:eMBBULLsensing}(a), where every third slot is reserved for low latency transmission. However, as also seen from the figure, if the low-latency traffic requests are random and intermittent, a portion of these reserved resources would be wasted (see, e.g., \cite{popovski20185g}). Conversely, upon too many arrivals, some of them may not be served as the latency deadline is exceeded before a dedicated resource becomes available, such as for the arrival \#3 on Fig.~\ref{fig:eMBBULLsensing}(a). To mitigate this problem and enhance the spectral efficiency, the transmission of both types of traffic can be jointly designed and adapted to current traffic conditions by leveraging SE plane functionalities. For example, semantic redundancy may arise from the transmission of multimedia signals that are never used since they carry details of the scene that the user does not interact with. SE plane functions can detect and remove this redundancy, hence freeing resources for low-latency traffic. This is illustrated in Fig.~\ref{fig:eMBBULLsensing}, where the packets denoted as ``broadband irrelevant'' in the top part are filtered by the SE plane in the bottom figure, enabling low-latency access for all packets.

\subsection{Semantic Control for Integrated Communication and Radar Sensing}

With the growing presence of mmWave and the advent of THz communications, there is an increasing focus on approaches to repurpose communication resources for radar sensing, or more generally for sub-millimeter-wave sensing and processing \cite{radar2}. A key problem is how to enable the control of radio resources for sensing without affecting the operation of the communication protocol. A standardized interface between the SE plane and the lower layers of the protocol stack would address this issue, making it possible to seamlessly integrate solutions by different vendors of both radar sensing and communication systems.

As an example, the MAC frame could include some blank slots, with suitable time and bandwidth guard bands, that may be directly controlled by the sensing applications run by the SE plane. Alternatively, the physical layer may include new modulation and coding schemes that offer degrees of freedom on the transmitted waveform that can be used for sensing applications. This idea generalizes the principle of protocol coding introduced in \cite{popovski2010protocol}, whereby degrees of freedom in the transmission protocol, e.g., the ordering of packets, can be leveraged to encode additional information on the communication signals.

\begin{figure}[t!]
\centering
\includegraphics{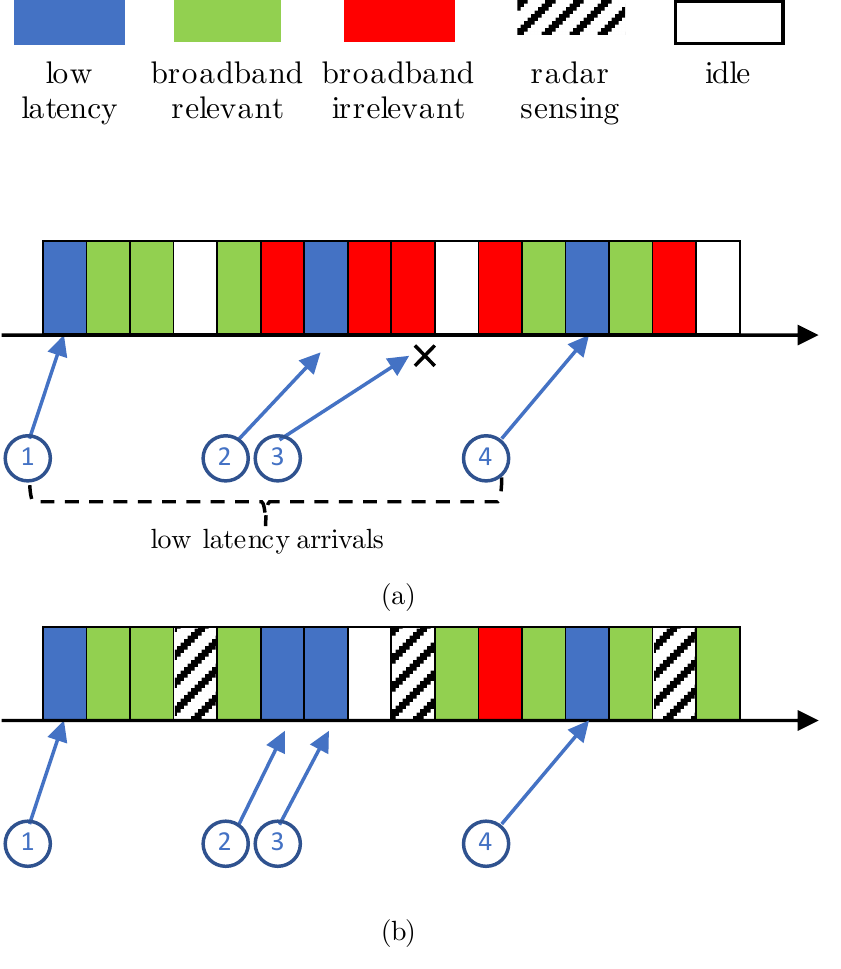}%
\caption{Illustration of multiplexing of low-latency wireless, broadband, and radio sensing in a MAC frame. (a) Low-latency resources are reserved periodically every three slots, and, if a low-latency packet needs to wait more than three slots, it is dropped; (b) SE filtering allows broadband packets carrying irrelevant information to be dropped, freeing resources for low-latency packets or other applications such as radar sensing. }
\label{fig:eMBBULLsensing}
\end{figure}

SE filtering and control can be integrated in a feedback cycle to reinforce one another. To illustrate this point, consider the examples of SE filtering and control covered so far in this section. As illustrated in Fig.~\ref{fig:eMBBULLsensing}, SE filtering of an AR application can identify idle resources and free them by dropping irrelevant data, so that they can be re-purposed for radar sensing. In turn, sensing signals can be used to predict users' physical interactions in the AR application; and thus, enable enhanced SE filtering of irrelevant broadband packets and reduction in the protocol overhead. 

%Being noisy, this prediction is imperfect (there is still one irrelevant broadband packet in Fig.~\ref{fig:eMBBULLsensing}(b), but it can still bring significant advantages. 

%the radar/radio sensing. This represents an input to the semantic plane at the physical layer. Based on the radio sensing and detection of the dynamics of the users, the communication system can anticipate gestures and preemptively allocate the communication resources. It is  justifiable to use some communication resources for radar due to the following: (1) radio sensing is a valuable input to the overall ML engine that attempts to predict the allocation of the broadband and low-latency communication resources. (2) if the application overhead is correctly anticipated, then with the re-purposing of the radio resources for radar use we decrease the application overhead.   

\subsection{Semantic Control for Physical-layer Computing}

In distributed computing over a wireless network, a number of devices hold local data, and the end goal is for a receiver to compute a function of the data distributed across these devices. Examples of functions include aggregated statistics via sums or extremal values. Function computation can be considered a semantic task that is handled by current communication protocols solely at the application layer. Accordingly, each device would communicate its separate statistic to the receiver, which would then compute the desired function at the application layer. In contrast, an SE control interface to the lower layers of the communication protocol can implement over-the-air computing techniques \cite{2019arXiv190100844A}, hence significantly enhancing the spectral efficiency of the system.

To provide a concrete example, consider implementing a supervised learning algorithm across wireless devices, each with its own local dataset. Current solutions for such tasks require devices to offload their local datasets to a cloud server, where a powerful learning algorithm, e.g., a deep neural network, is executed on the accumulated dataset, and the learned model is shared with the devices. However, privacy concerns, or the sheer size of the dataset, may prevent this centralized solution. Alternatively, in federated learning \cite{FL:KonecnyMRR16} the devices can be orchestrated by a parameter server, e.g., a base station or another wireless device, to learn a model in a distributed manner. At each iteration of the algorithm, each device updates and sends its local model or gradient estimate, and the goal of the parameter server is to compute the average of these to update the global model, which is multicast back to the devices. 

\begin{figure}[!t]
\centering
\includegraphics[width=\columnwidth]{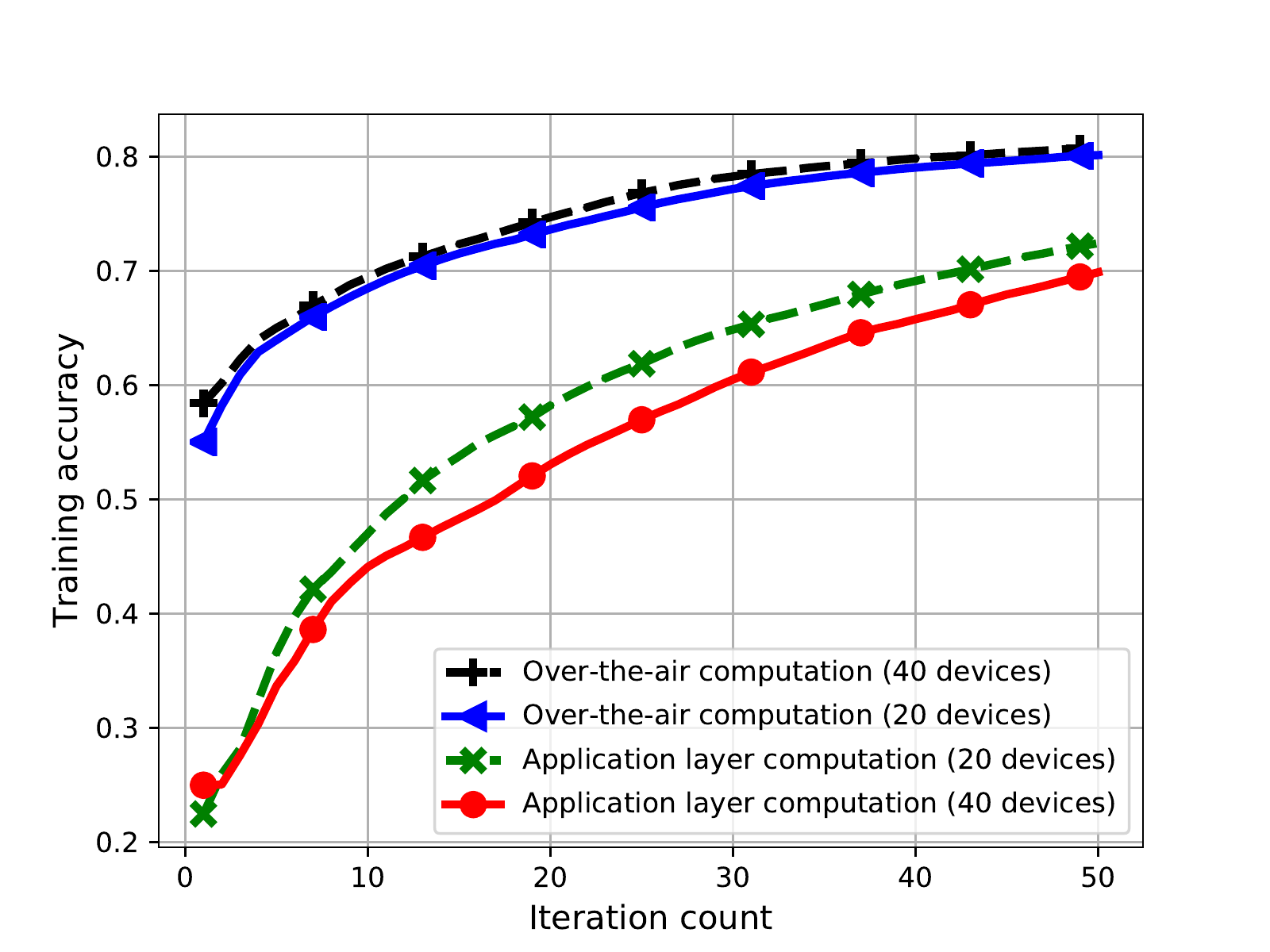}%

\caption{Comparison of the training accuracy achieved by application layer computation vs. over-the-air computation for federated MNIST classification across wireless devices.}
\label{fig_MNIST}
\end{figure}

To provide an illustration, in Fig.~\ref{fig_MNIST}, the accuracy of a single-layer neural network trained for the classical MNIST digit classification problem by means of gradient averaging is shown versus the number of training iterations \cite{2019arXiv190100844A}. We compare the performance of the standard method that computes the average at the application layer, and of over-the-air computation enabled through an SE plane. In the latter case, the average gradient is directly estimated from the received signal thanks to the superposition of wireless signals at the physical layer. We observe that over-the-air computation results in a significantly higher training accuracy, with a gain that increases with the number of devices. On the contrary, the accuracy of application layer computation decreases with the number of devices, due to the decreasing amount of spectral resources per device to send their local estimates separately.

%, which means less channel resources per device, and hence, less accurate transmission of local estimates to the parameter server, resulting in a weaker learning performance. 

\subsection{Other Examples of SE Filtering and Control}
Having detailed a few applications of SE filtering and control, here we briefly list a few additional use cases.

\noindent $\bullet$ \emph{Physical-layer provenance filtering}: Validate data provenance using radio fingerprinting or location-based authentication through baseband signal processing, while removing the need for higher-layer operations (see, e.g., \cite{xu2016device});
%\item \emph{Physical-layer computing}: Compute aggregated statistics by means of physical-layer computing, hence reducing semantic and protocol overheads by avoiding the need to transmit separate data points \cite{nazer2011compute, compfed18, 2019arXiv190100844A}.

\noindent $\bullet$ \emph{Physical-layer remote radio control}: Allow access to physical resources, such as smart locks, via radio signatures controlled directly by the SE plane through trust mechanisms such as smart contracts. As an example, once rent is paid, a smart contract controls the physical layer transmission of a radio signature to open a smart lock \cite{roush2018twelve};
%\item \emph{Physical-layer integration of mmwave/THz radar and communication}: Use the same radio interface for both radar and communications~\cite{choi2016millimeter}, hence increasing efficiency and enabling new applications such as gesture-based interfaces. 

\noindent $\bullet$ \emph{Layer-2 retransmission control}: Retransmit only data that is expected to be still relevant on the basis of the internal state of ML/AI models by means of SE filtering (see, e.g., \cite{zhu2018towards});

\noindent $\bullet$ \emph{Network-layer traffic-based routing}: Classify traffic on the basis of its network-level traces and configure routers accordingly to filter out protocol overhead;

\noindent $\bullet$ \emph{Transport-layer semantic-based access and congestion control}: Inject data in the network only when relevant and trusted, hence reducing protocol overhead. A notable example is given by sensors and actuators that operate on the basis of input from local or cloud-based predictive models. Accordingly, based on past data collected by sensors and actuators, an ML/AI module can instruct a sensor on how to carry out sampling and transmission and can inform an actuator of any change in its control module only when the local model is outdated;

%predictive models. An example of the decrease in the semantic overhead can be seen in a system with sensors and actuators connected through a cloud/core network. There is an ML module in the cloud that learns from the past data and interactions among the sensors, actuators and the environment. This module creates passes on (and updates) models to the sensors and the actuators, to be integrated in their semantic planes. Based on the input from the semantic plane: \emph{(i)} a sensor starts to become more selective which data will it transmit; \emph{(ii)} an actuator makes predictions about its actions and uses the data from the sensors only to increase the belief in the actions. If the sensing data is excessively delayed, then the actuator takes action solely based on the prediction.
\noindent $\bullet$ \emph{Application-level aggregation for DLT transactions}: Aggregate transactions from multiple applications running on a device or a network node in a single verifiable unit (e.g., block) prior to communications within a distributed ledger for, e.g., financial transactions or smart contracts, hence reducing the protocol overhead;

%A large number of applications and services will rely on smart contracts and transactions carried through a DLT or blockchains. The semantic plane will have the role of  and in this way decrease the amount of traffic generated by DLT. This will have a direct impact on 
\noindent $\bullet$ \emph{Intent-based networking}: Upon receiving users' instructions via natural language or visual interfaces, the network automatically reconfigure itself at all layers of the protocol stack in order to carry out the described task in the most efficient manner.

%Note: Do we need a new semantic plane or could we implement it exploiting SDN capabilities? E.g. could we do it via a software update of SDN routers? If that's case, we could demonstrate it on real-world hardware via a proof-of-concept. In my view, some functionalities, particularly those related to layer 3 can be directly implemented using SDN. In a way, what we are proposing is a generalization of the SDN approach to all layers of the protocol stack.

\section{SE Functionalities in 5G and 3GPP}
The proposed SE plane has precursors in the architectures that are currently under discussion within 3GPP. This section provides a brief overview of the corresponding state of the art.

%\subsection{5G Network Data Analytics}

The 5G SBA is designed to support the interaction of control-plane NFs. The Network Data Analytics Function (NWDAF) \cite{3GPPTS23.501} \cite{3GPPTS23.288}  is one of the NFs in the 5G core system architecture. NWDAF is used for data collection and analytics, and  utilizes the service-based model to communicate with other NFs and the Operation, Administration and Maintenance (OAM) system. The purpose of standardizing NWDAF,  rather than relying on proprietary implementations, is to allow mobile network operators to handle analytics for NFs and third-party services. For example, NWDAF can be used to extract data from a NF, to derive statistical information about the past, or predictive information about future events, and to provide this information to other NFs or to the OAM. These NFs or the OAM system can then take analytics-driven decisions to implement, e.g., proprietary self-organizing network tools. We emphasize that 3GPP does not standardize the data analytics algorithms, but only the interfaces and related procedures that support inter-NF communications, such as subscribe/unsubscribe procedures, as well as the type of information NWDAF can examine such as user equipment (UE) mobility and user plane congestion.
%software-based functionalities and services such that various network capabilities can be flexibly deployed and scaled as needed [1]\color{red} Fede and Onur, please update references in this section \color{black}. The control plane NFs can interact with each other using a service-based model; in other words, a given control plane NF can enable authorized NFs to access its services. 
To be more specific, in the 5G New Radio (NR) standard, as defined by the current Release 16 \cite{3GPPTS23.288} and illustrated in Fig. 3, NWDAF interacts with and receives analytics-related information from different entities in the network, including the Access and Mobility Function (AMF), the Session Management Function (SMF), the Network Exposure Function (NEF), and the Policy Control Function (PCF). AMF allows NWDAF to access UE and base station location and mobility analytics; SMF allows NWDAF to receive session-related analytics; NEF allows NWDAF to access third party-related (e.g. YouTube, Facebook, etc.) service analytics; whereas PCF allows NWDAF to obtain policy-related information.

\begin{figure}[!t]
\centering
\includegraphics[width=8.3cm]{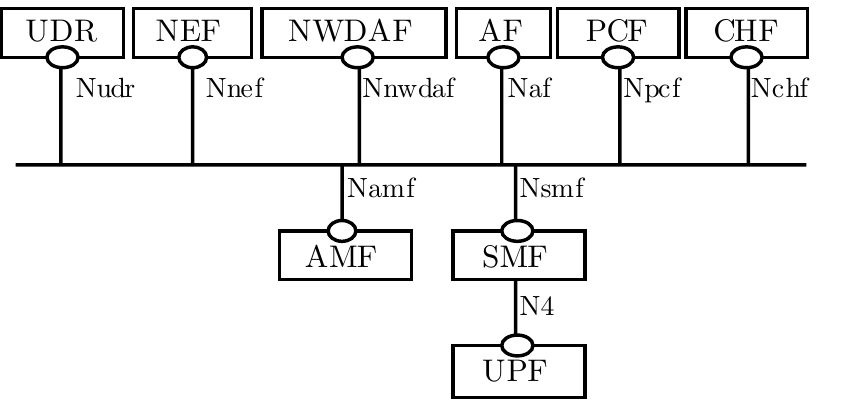}%
\caption{State of the art for ML/AI interfaces in the core network of 5G networks: Interaction of NWDAF with other network functions.}
\label{fig3}
\end{figure}

More recently, as part of Release 16 discussions, the 3GPP RAN group has started a study item on ``RAN-centric data collection and utilization for NR'', with the aim of broadening the set of standard-based solutions for enhanced data collection and utilization capabilities \cite{3GPPTR37.816}. The study item, which is currently in its early stages, has the objective to investigate features that could enable interaction of the RAN and the enhanced NDWAF blocks. It primarily aims to identify relevant data, such as Layer-1 and Layer-2 measurement quantities, and to investigate procedures that collect and utilize that data for automated network functionalities. Examples of applications under investigation include mobility optimization, random access channel optimization, load sharing/balancing related optimization, coverage and capacity optimization, and minimization of drive testing.

\section{Discussion and Conclusions}

We conclude this paper with a brief discussion about challenges and implications of the proposed introduction of a SE plane.

We highlight the following two main key challenges towards the implementation of the SE plane concept.
\begin{itemize}
    \item \emph{Cross-Layer Recovery Mechanisms}. The direct control of lower layers by the SE plane risks breaking mechanisms that have been carefully designed based on a hierarchy between different layers. As an example, the semantic layer may apply a filtering decision at Layer 2, that can have an implication on TCP retransmissions and reordering at Layer 4. Design of proper cross-layer mechanisms and interfaces should address this important technical problem.
    \item \emph{Security.} Data plane security today is mainly based on application, transport, and network layer mechanisms. In that context, it is not clear what are the security implications of exposing Application Programming Interfaces (APIs) below the network layer. For example, is there a risk in opening some new backdoors? Possible solutions to this problem could handle security directly at the SE plane and could be embedded in the API exposed to users, sensors, and actuators. For example, the SE plane may carry out physical layer-based authentication prior to granting access to some functionalities. 
\end{itemize}

The standardization of the SE plane architecture introduced in this paper would by and large revolve around the definition of effective interfaces between communication layers and application programming interfaces towards users, actuators, and devices.  
The availability of well-defined software interfaces for SE filtering and control may invalidate the current ``next-G paradigm'' for mobile wireless evolution and standardization, ushering in an era of continuous ``open-source'' improvements and extensions.  In this profoundly different ecosystem, defined by the separation between hardware and software and by standardised data analytics and control interfaces, the availability of open-source software running on general purpose hardware and the maturity of software defined radio solutions will lower the barriers for new entrants and enable faster innovation. 

\ifCLASSOPTIONcaptionsoff
  \newpage
\fi



%\bibliographystyle{IEEEtran}
%\begin{thebibliography}{1}
%\end{thebibliography}

\bibliographystyle{IEEEtran}
\end{document}